\documentclass[final,5p,times,twocolumn]{elsarticle} 

\usepackage{graphicx}

\usepackage{amssymb}

\journal{Nuclear Instruments and Methods A}

\begin{document}

\begin{frontmatter}



\title{Silica aerogel radiator for use in the A-RICH system utilized in the Belle II experiment}


\author[A,B]{Makoto Tabata\corref{cor1}}
\ead{makoto@hepburn.s.chiba-u.ac.jp}
\cortext[cor1]{Corresponding author.} 
\author[C]{Ichiro Adachi}
\author[D]{Nao Hamada} 
\author[C]{Koji Hara}
\author[E]{Toru Iijima}
\author[F]{Shuichi Iwata}
\author[F]{Hidekazu Kakuno}
\author[B]{Hideyuki Kawai}
\author[G,H]{Samo Korpar}
\author[I,H]{Peter Kri\v{z}an}
\author[F]{Tetsuro Kumita}
\author[C]{Shohei Nishida}
\author[D]{Satoru Ogawa}
\author[H]{Rok Pestotnik}
\author[H]{Luka \v{S}antelj}
\author[H]{Andrej Seljak}
\author[F]{Takayuki Sumiyoshi}
\author[H]{Elvedin Tahirovi\'{c}}
\author[F]{Keisuke Yoshida}
\author[J]{Yosuke Yusa}
\address[A]{Institute of Space and Astronautical Science (ISAS), Japan Aerospace Exploration Agency (JAXA), Sagamihara, Japan}
\address[B]{Department of Physics, Chiba University, Chiba, Japan}
\address[C]{Institute of Particle and Nuclear Studies (IPNS), High Energy Accelerator Research Organization (KEK), Tsukuba, Japan}
\address[D]{Department of Physics, Toho University, Funabashi, Japan}
\address[E]{Kobayashi--Maskawa Institute for the Origin of Particles and the Universe, Nagoya University, Nagoya, Japan}
\address[F]{Department of Physics, Tokyo Metropolitan University, Hachioji, Japan}
\address[G]{Faculty of Chemistry and Chemical Engineering, University of Maribor, Maribor, Slovenia}
\address[H]{Experimental High Energy Physics Department, Jo\v{z}ef Stefan Institute, Ljubljana, Slovenia}
\address[I]{Faculty of Mathematics and Physics, University of Ljubljana, Ljubljana, Slovenia}
\address[J]{Department of Physics, Niigata University, Niigata, Japan}

\begin{abstract}

This paper presents recent progress in the development and mass production of large-area hydrophobic silica aerogels for use as radiators in the aerogel-based ring-imaging Cherenkov (A-RICH) counter, which will be installed in the forward end cap of the Belle II detector. The proximity-focusing A-RICH system is especially designed to identify charged kaons and pions. The refractive index of the installed aerogel Cherenkov radiators is approximately 1.05, and we aim for a separation capability exceeding 4$\sigma $ at momenta up to 4 GeV/$c$. Large-area aerogel tiles (over 18 $\times $ 18 $\times $ 2 cm$^3$) were first fabricated in test productions by pin drying in addition to conventional methods. We proposed to fill the large end-cap region (area 3.5 m$^2$) with 124 water-jet-trimmed fan-shaped dual-layer-focusing aerogel combinations of different refractive indices (1.045 and 1.055). Guided by the test production results, we decided to manufacture aerogels by the conventional method and are currently proceeding with mass production. In an electron beam test undertaken at the DESY, we confirmed that the $K$/$\pi $ separation capability of a prototype A-RICH counter exceeded 4$\sigma $ at 4 GeV/$c$.

\end{abstract}


\end{frontmatter}

\section{Introduction}

\begin{figure}[t]
\centerline{
\includegraphics[width=0.95\columnwidth]{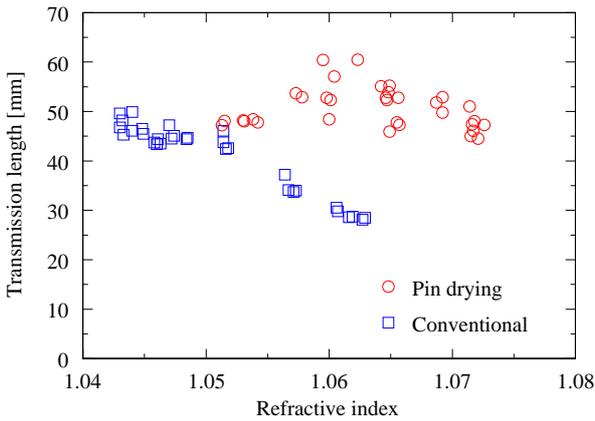}}
\caption{Transmission length at 400 nm as a function of refractive index. The transmission length was calculated from the aerogel thickness and transmittance measured with a spectrophotometer \cite{14}. The refractive index was measured at the tile corners by the Fraunhofer method with a 405-nm semiconductor laser \cite{14}. The plot compares aerogel samples smaller than (10 $\times $ 10 $\times $ 2) cm$^3$ produced by the conventional (squares) and pin-drying (circles) methods.}
\label{fig1}
\end{figure}

We are developing an aerogel ring-imaging Cherenkov (A-RICH) counter \cite{1} as one of the particle identification devices for the Belle II experiment \cite{2}. Belle II will involve the SuperKEKB electron--positron collider, under upgrade at KEK, Japan. This super $B$-factory experiment, investigating flavor physics and precision measurements of $CP$ violations in the intensity frontier, will search for physics beyond the standard model of particle physics. Because the installation space of the forward end cap of the Belle II detector is limited, the A-RICH system was designed as a proximity-focusing RICH counter. The A-RICH system is an upgrade of threshold aerogel Cherenkov counters \cite{3} installed in the previous Belle spectrometer \cite{4}.

To efficiently separate kaons from pions at momenta ($p$) up to 4 GeV/$c$ (our goal is to exceed 4$\sigma $ $K$/$\pi $ separation capability), we require a highly transparent Cherenkov radiator with a refractive index $n$ of approximately 1.05. To this end, we have been developing high-refractive index, high-quality hydrophobic silica aerogels since the early 2000s \cite{5,6,7,8}. Together with Hamamatsu Photonics K.K., Japan, we have also been developing 144-ch multi-anode hybrid avalanche photodetectors (HAPD) \cite{9} as position-sensitive Cherenkov light sensors using dedicated readout electronics (ASIC) \cite{10}.

Given the limited expansion distance between the aerogel upstream surface and photodetector surface (20 cm in our case), we first proposed and demonstrated a multilayer-focusing radiator scheme using aerogels with different refractive indices \cite{11}. This thick aerogel combination aimed to increase the number of detected photoelectrons without degrading the resolution of the Cherenkov angle. In previous beam tests, the $K$/$\pi $ separation capability of a prototype A-RICH counter exceeded 5$\sigma $ at 4 GeV/$c$ \cite{12}. Such high separation capability was achieved using highly transparent (but small size) aerogels developed by the pin-drying method described in the next section.

Silica aerogel, an amorphous, highly porous solid of silicon dioxide (SiO$_2$), has been widely used as a Cherenkov radiator because of its tunable, intermediate refractive index and optical transparency. Its refractive index is related to its bulk density, which depends on the silica--air volume ratio (typically 1:9) tuned in the production process. The typical length scale of three-dimensional networks of silica particles (revealed by our scanning electron microscopy observations) is on the order of 10 nm. Hence, the transmission of Cherenkov photons in an aerogel is described by Rayleigh scattering. Although aerogels are generally transparent, their transparency strongly depends on production technique.

\section{Production methods}

Our aerogels are produced in two ways: conventional and pin-drying methods. Our conventional method, developed at KEK in the 1990s \cite{13}, was modernized in the mid-2000s by introducing a new solvent in the wet-gel synthesis process [5]. The conventional method, which requires approximately one month, adopts a simple wet-gel synthesis procedure (sol--gel step). The refractive indices of the resulting aerogels reach $n$ = 1.11 \cite{14}. To suppress age-related degradation caused by moisture absorption, our aerogels are rendered hydrophobic \cite{14,15}. In the final step, the wet gels are dried by the supercritical carbon dioxide extraction method.

Our recently developed pin-drying method \cite{16} produces more transparent aerogels (with $n \sim $1.06) than the conventional method. Originally, this novel method was developed to yield ultrahigh-refractive-index aerogels (up to $n$ = 1.26) at Chiba University, Japan \cite{17}. In this method, individual wet gels (originally designed with $n \sim $1.05) are subjected to a special treatment, intermediate between the wet-gel synthesis and hydrophobic treatment. Specifically, to increase its silica density, the wet gel is shrunk by partial drying in a semi-sealed container punctured with pinholes to suppress cracking (pin-drying process). The refractive index of the aerogels is determined by two factors: shrinkage extent of the wet gel and the volume ratio of chemicals used in the wet-gel synthesis. By contrast, the refractive index is controlled only by the wet-gel synthesis process in the conventional method. Potential disadvantages of the pin-drying method include the time consumed in the pin-drying process (approximately two months), difficulties in obtaining uniform tile density, and cracking in the supercritical drying phase. However, the method achieves long transmission length ($\Lambda _{\rm T}$) \cite{8}, as shown in Fig. \ref{fig1}.

\section{Tiling in the end cap}

When filling the large (3.5 m$^2$) end-cap region, tile boundaries must be reduced by minimizing the total number of aerogel tiles (i.e., by maximizing aerogel dimensions). This is important because unintended scattering at the boundaries of adjacent aerogel tiles reduces the number of detected photoelectrons. Meanwhile, the tile dimensions should also be realistic for production. Considering the capacity of our autoclave used in the supercritical drying phase, installed at Mohri Oil Mill Co., Ltd., Japan, we planned to manufacture aerogel tiles with dimensions (18 $\times $ 18 $\times $ 2) cm$^3$.

\begin{figure}[t]
\centerline{
\includegraphics[width=0.8\columnwidth]{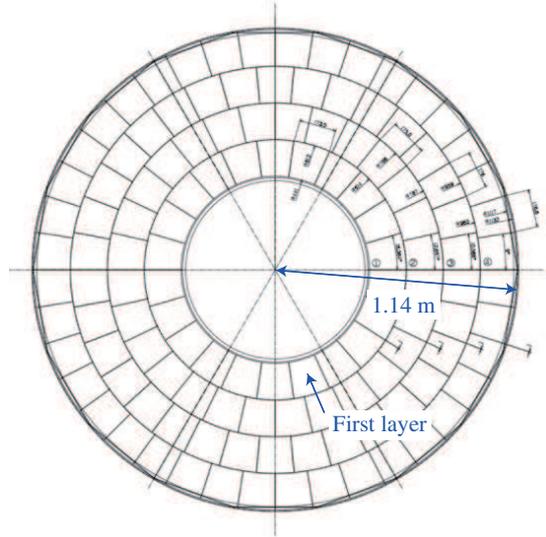}}
\caption{CAD drawing of the planned A-RICH radiator tiling scheme in the cylindrical forward end cap of the Belle II detector.}
\label{fig2}
\end{figure}

To simplify aerogel production, we propose to fill  the cylindrical end-cap region with 124 segmented dual-layer-focusing aerogel combinations (a total of 248 tiles) with different refractive indices. Fig. \ref{fig2} shows a CAD drawing of the planned aerogel radiator tiling scheme. Each 2-cm-thick aerogel layer (dual layer thickness = 4 cm) is separately fabricated and will be stacked during the installation of the A-RICH system.

\begin{figure}[htb]
\centerline{
\includegraphics[width=0.95\columnwidth]{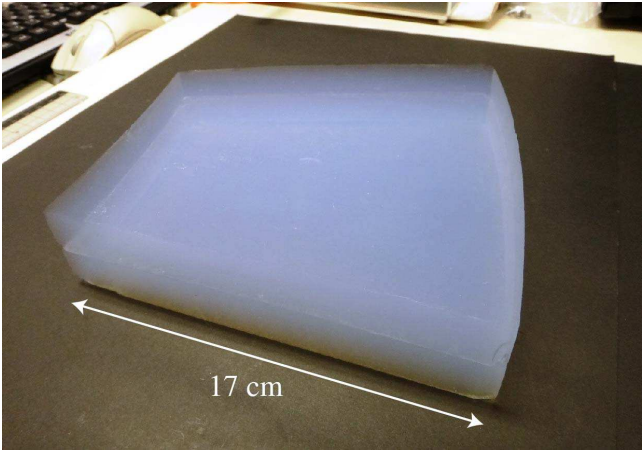}}
\caption{Fan-shaped water-jet-machined aerogels to be installed in the first layer of the end cap. These were cut from (18 $\times $ 18 $\times $ 2) cm$^3$ tiles and stacked to form a focusing combination.}
\label{fig3}
\end{figure}

Prepared aerogel tiles are cut into fan shapes that fit their layer (concentric layers 1--4, counting from the inside of the end cap; see Fig. \ref{fig2}) by a water jet cutter. Water jet trimming best exploits the hydrophobic features of our aerogels. Fig. \ref{fig3} shows water-jet-machined aerogels cut from (18 $\times $ 18 $\times $ 2) cm$^3$ test production tiles fabricated by the conventional method.  The integrity of the optical parameters was preserved after cutting, and the error in the tile dimensions was below 1\%.

In addition, to increase the number of Cherenkov photons that reach the photodetectors without Rayleigh scattering, the aerogels should be maximally transparent. At $n$ = 1.045 for upstream aerogels, sufficient transparency is achieved by the conventional method; however, transparency generally decreases with increasing refractive index. On the other hand, downstream aerogels with $n \sim $1.055 produced by the pin-drying method are more transparent than those obtained from the conventional method. Therefore, the upstream aerogels will be manufactured by the conventional method, while the downstream aerogels can be produced by either method. Fabricating the downstream ($n$ = 1.055) aerogels by the conventional method is a promising option, because large aerogels can be obtained with a crack-free yield exceeding 80\% \cite{18}. Alternatively, the pin-drying method is suitable in cases in which the transparency of the downstream aerogels is particularly important for detector performance; that is, where all emitted Cherenkov photons must pass through the downstream layer to reach the photodetectors. Of special concern in the pin-drying method is obtaining large aerogels with low crack-free yield.


\section{Test production results of large-area pin-dried tiles}

For the actual detector, we require approximately 350 large-area aerogel tiles (including spares) with no cracking. For large, high-refractive-index aerogels, cracking in the supercritical drying phase is a critical issue. The total number of wet gels that must be synthesized and processed depends on the crack-free yield in the final phase. In test productions by the conventional method, we established a technique for fabricating large-area tiles with a sufficiently high crack-free yield.

However, cracking remains problematic in large-area aerogels test produced by the pin-drying method. Cracking is more critical in the pin-drying method than in the conventional method, even during small aerogel production. However, in 2012, the first intact (17 $\times $ 17 $\times $ 2) cm$^3$ tile with $n \sim $1.059 was obtained \cite{19}. To date, these are the maximum crack-free tile dimensions obtainable by the pin-drying method. In 2013, immediately prior to mass production, a two-batch (52 tiles) test production of (19 $\times $ 19 $\times $ 2) cm$^3$ aerogels with $n$ = 1.055--1.060 was achieved by the pin-drying method, although all of the attained samples were cracked.

From the above test production results of large-area aerogels, we have conclusively decided to mass produce both the upstream ($n$ = 1.045) and downstream ($n$ = 1.055) aerogels by the conventional method. We consider that intact aerogels are essential to ensure the safe handling of aerogels with no fragmentation at the radiator installation stage. Moreover, in the cracking plane, Cherenkov photons are scattered or reflected, which measurably decreases the transmittance of the aerogels. The optical parameters of the recently produced aerogels were measured and found to be suitable even in large tiles \cite{19}. The measured transmission lengths were 40 mm--50 mm and 35 mm for aerogels with $n$ = 1.045 and 1.055, respectively. An X-ray technique \cite{20} revealed that the density nonuniformity of the tiles in the transverse planar direction was below $\pm $1\%.

\section{Beam test of a prototype A-RICH system}

\begin{figure}[t]
\centerline{
\includegraphics[width=0.95\columnwidth]{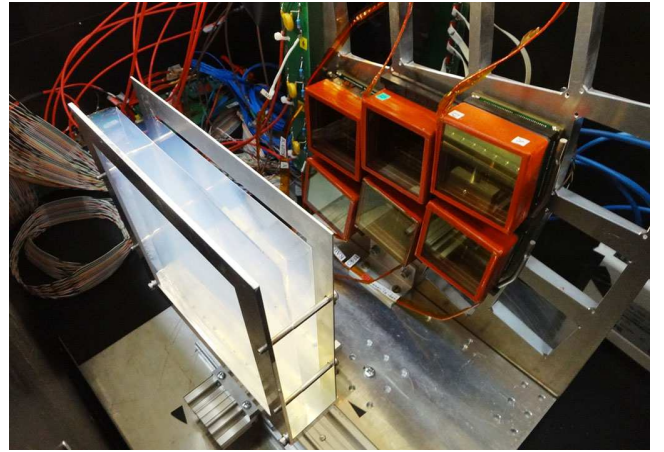}}
\caption{Experimental set up of the prototype A-RICH counter in the light-shielded box.}
\label{fig4}
\end{figure}

In May 2013, the two above-described options for large-area aerogel combinations were evaluated in a beam test undertaken at the DESY. In the test, a prototype A-RICH counter was exposed to a $p$ = 5 GeV/$c$ electron beam in the T24 beam area of the synchrotron. Spatial nonuniformity in the density of pin-dried aerogels has been recognized by X-ray absorption measurements \cite{19,20}. Therefore, we were particularly interested in whether refractive index nonuniformities (gradients) in the transverse planar direction of monolithic aerogel tiles would degrade detector performance. Independent of refractive index nonuniformity, the laser Fraunhofer method tags the aerogels with the refractive index measured at the tile corners ($n_{\rm {tag}}$). The previous test \cite{12} targeted pin-dried aerogels with $n_{\rm {tag}}$ = 1.054 and 1.065 in the upstream and downstream layers, respectively. These are small aerogels with dimensions of (9 $\times $ 9 $\times $ 2) cm$^3$. Unlike this study, the previous study paid little attention to refractive index nonuniformity.

The prototype A-RICH counter (as a component of the real-size detector) was constructed by correctly placing the system components (aerogel radiators, prototype HAPDs, and front-end electronics close to the final design) in the 20-cm expansion distance between the aerogel surface and HAPDs in a light-shielded box. The setup is shown in Fig. \ref{fig4}. Six HAPDs, separated by gaps, were arranged into a 2 $\times $ 3 array, mimicking the layout of the real detector. The pixel size and averaged quantum efficiency of the HAPDs were (4.9 $\times $ 4.9) mm$^2$ and 25\%, respectively. The charged beam was tracked by four multiwire proportional chambers affixed to the upstream and downstream sides of the light-shielded box. Trigger signals were generated by two scintillation counters placed at the upstream and downstream extremities of the beam area.

In a preliminary analysis of the data obtained from aerogel study runs, we obtained $N_{\rm p.e.}$ = 6.6 and $\sigma_{\rm {\theta}}$ = 14.5 mrad, implying a $K$/$\pi $ separation capability of 4.1$\sigma $. Here a focusing aerogel combination of $n$ = 1.0464 ($\Lambda _{\rm T}$= 42 mm) and 1.0548 (34 mm) was constructed from tiles produced by the conventional method. $N_{\rm p.e.}$ is the averaged number of detected photoelectrons, and $\sigma_{\rm {\theta}}$ is the resolution of the observed Cherenkov angle distribution per single photon. The performance of the A-RICH detector was na\"{i}vely estimated by the following equation:
\[ \frac{\Delta \theta_{\rm C}\sqrt{N_{\rm p.e.}}}{\sigma_{\rm \theta}}, \]
where $\Delta \theta_{\rm C}$ ($\sim $23 mrad) is the difference in Cherenkov angles between $K$ and $\pi $ at $n$ = 1.05 and $p$ = 4 GeV/$c$. The aerogels incorporated into the prototype detector were the water-jet-trimmed tiles shown in Fig. \ref{fig3}. In this run, the electron beam impacted perpendicular to the default aerogel position (around the center of the tiles). This aerogel combination, selected for our mass production, approximately satisfied our detector performance requirements.

\begin{figure}[htb]
\centerline{
\includegraphics[width=0.95\columnwidth]{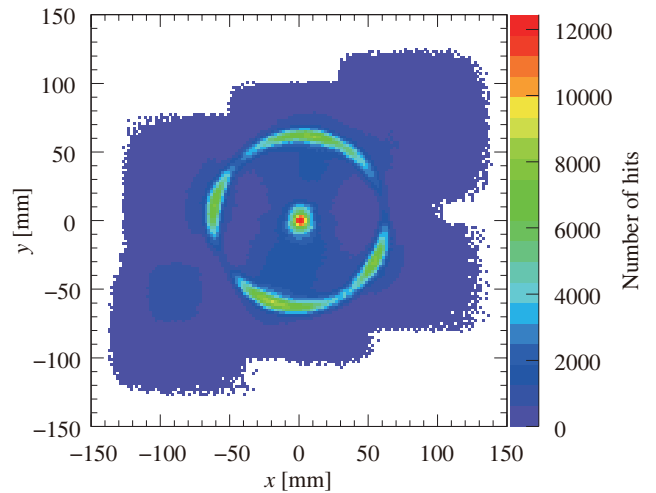}}
\caption{Cherenkov ring observed in the conventional and pin-dried aerogel combination. During the run, the distribution of striking Cherenkov photons in each event was accumulated on the photodetection plane. The difference between the beam and HAPD hit position is plotted for each hit.}
\label{fig5}
\end{figure}

An alternative focusing aerogel combination, namely, $n$ = 1.0467 ($\Lambda _{\rm T}$ = 47 mm) and $n_{\rm {tag}}$ = 1.0592 (59 mm), constructed from tiles produced by the conventional and pin-drying method, respectively, yielded a superior $K$/$\pi $ separation capability of 4.5$\sigma $. In this combination, $N_{\rm p.e.}$ = 8.6 and $\sigma_{\rm {\theta}}$ = 15.0 mrad. Accumulated clear Cherenkov rings are shown in Fig. \ref{fig5}. In the pin-dried aerogels, the tagged refractive index at the tile corners will be larger than the refractive index averaged across the tile, possibly reflecting a density gradient \cite{19,20}. We irradiated the tile center, corner, and midway point (between center and corner) with the beam. Ongoing analysis indicated that refractive index nonuniformity does not significantly affect detector performance; i.e., the spatial variation in the mean and resolution of the Cherenkov angle distribution appears to be within our requirements and is acceptable. We conclude that detector performance could be primarily improved by detecting more Cherenkov photons using highly transparent pin-dried aerogels.

\section{Status of mass production}

In September 2013, collaborating with the Japan Fine Ceramics Center and Mohri Oil Mill Co., Ltd, we began the mass production of aerogels for the actual A-RICH system. By the end of 2013, 100 tiles were sequentially delivered to KEK. In parallel, we characterized the aerogels by visual checking and measurements of refractive index, transmission length, and density. Thus far, our crack-free yield exceeds 90\% (outperforming our target of 80\%), and we have measured the optimal transparency of tiles produced by the conventional method. Mass production will be completed by the end of May, 2014.

\section{Conclusion}

As part of the Belle detector upgrade program, we are investigating an A-RICH counter to be installed in the forward end cap as a particle identification system. We introduced conventional and pin-drying methods for producing high-quality hydrophobic silica aerogels intended for use as Cherenkov radiators. We proposed an aerogel tiling scheme in the large-area end cap of the Belle II detector. We then studied a technique for producing large aerogel tiles and examined the crack-free yield of samples produced by both the conventional and pin-drying methods. Based on the test production results, we elected to mass produce both upstream ($n$ = 1.045) and downstream ($n$ = 1.055) aerogels by the conventional method. Mass production and quality checking are now fully established. We constructed a prototype A-RICH counter from tiles test produced by the conventional method arrayed in a dual-layer-focusing radiator scheme and confirmed its utility in a test beam experiment at the DESY. We also confirmed the superior detector performance of a highly transparent pin-dried aerogel. The pin-drying process will be further developed at Chiba University, independent of the Belle II program.



\section*{Acknowledgments}

The authors are grateful to Japan Fine Ceramics Center and Mohri Oil Mill Co., Ltd. for their contributions to large aerogel production. We are also grateful to the Venture Business Laboratory at Chiba University for offering rooms for manufacturing aerogels. This study was partially supported by a Grant-in-Aid for Scientific Research (A) (No. 24244035) from the Japan Society for the Promotion of Science (JSPS). M. Tabata was supported in part by the Space Plasma Laboratory at ISAS, JAXA.


\begin{thebibliography}{00}

\bibitem{1}
R. Pestotnik, et al., Nucl. Instrum. Methods A 732 (2013) 371; S. Nishida, et al., Nucl. Instrum. Methods A, this issue.
\bibitem{2}
T. Abe, et al., Belle II Technical Design Report, KEK Rep. 2010-1, 2010.
\bibitem{3}
T. Sumiyoshi, et al., Nucl. Instrum. Methods A 433 (1999) 385; T. Iijima, et al., Nucl. Instrum. Methods A 453 (2000) 321.
\bibitem{4}
A. Abashian, et al. (Belle Collaboration), Nucl. Instrum. Methods A 479 (2002) 117; also see detector section in J. Brodzicka, et al., Prog. Theor. Exp. Phys. (2012) 04D001.
\bibitem{5}
I. Adachi, et al., Nucl. Instrum. Methods A 553 (2005) 146.
\bibitem{6}
I. Adachi, et al., Nucl. Instrum. Methods A 581 (2007) 415.
\bibitem{7}
I. Adachi, et al., Nucl. Instrum. Methods A 595 (2008) 180.
\bibitem{8}
I. Adachi, et al., Nucl. Instrum. Methods A 639 (2011) 222.
\bibitem{9}
S. Korpar, et al., Nucl. Instrum. Methods A, this issue.
\bibitem{10}
H. Kakuno, et al., Nucl. Instrum. Methods A, this issue.
\bibitem{11}
T. Iijima, et al., Nucl. Instrum. Methods A 548 (2005) 383.
\bibitem{12}
S. Shiizuka, et al., Nucl. Instrum. Methods A 628 (2011) 315; K. Hara, et al., in Conf. Rec. on IEEE Nucl. Sci. Symp. and Med. Imaging Conf., 2010, p. 415.
\bibitem{13}
I. Adachi, et al., Nucl. Instrum. Methods A 355 (1995) 390.
\bibitem{14}
M. Tabata, et al., Nucl. Instrum. Methods A 668 (2012) 64.
\bibitem{15}
H. Yokogawa and M. Yokoyama, J. Non-Cryst. Solids 186 (1995) 23.
\bibitem{16}
M. Tabata, et al., Phys. Proc. 37 (2012) 642.
\bibitem{17}
M. Tabata, et al., in: Conf. Rec. on IEEE Nucl. Sci. Symp. and Med. Imaging Conf., 2005, p. 816; M. Tabata, et al., Nucl. Instrum. Methods A 623 (2010) 339.
\bibitem{18}
M. Tabata, et al., in: Conf. Rec. on IEEE Nucl. Sci. Symp. and Med. Imaging Conf., 2013, NPO1-153.
\bibitem{19}
M. Tabata, et al., in: Conf. Rec. on IEEE Nucl. Sci. Symp. and Med. Imaging Conf., 2012, p. 458.
\bibitem{20}
M. Tabata, et al., IEEE Trans. Nucl. Sci. 59 (2012) 2506.

\end{thebibliography}
\end{document}